\documentclass[a4paper,12pt]{article}
\usepackage[top=4cm,bottom=4cm,left=3cm,right=3cm]{geometry}
\usepackage{graphicx}
\usepackage{amsmath, amsthm, amsfonts,multirow,float,moreverb}
\numberwithin{equation}{section}
\usepackage{authblk}
\usepackage{booktabs}
\usepackage[toc,page]{appendix}
\usepackage{setspace}
\usepackage[numbers]{natbib}
\usepackage[pdftex]{hyperref}
\title{Missing continuous outcomes under covariate dependent missingness in cluster randomised trials}
\author{Anower Hossain, Karla Diaz-Ordaz and Jonathan W. Bartlett}
\affil[]{\small Department of Medical Statistics\\ London School of Hygiene and Tropical Medicine (LSHTM)}
\date{}

\begin{document}
\maketitle

\begin{abstract}
Attrition is a common occurrence in cluster randomised trials (CRTs) which leads to missing outcome data. Two approaches for analysing such trials are cluster-level analysis and individual-level analysis. This paper compares the performance of unadjusted cluster-level analysis, baseline covariate adjusted cluster-level analysis and linear mixed model (LMM) analysis, under baseline covariate dependent missingness (CDM) in continuous outcomes, in terms of bias, average estimated standard error and coverage probability.  The methods of complete case analysis (CCA) and multiple imputation (MI) are used to handle the missing outcome data. Four possible scenarios are considered depending  on whether the missingness mechanisms and covariate effects on outcome are the same or different in the two intervention groups. We show that both unadjusted cluster-level analysis and baseline covariate adjusted cluster-level analysis give unbiased estimates of the intervention effect only if both intervention groups have the same missingness mechanisms and the same covariate effects, which is arguably unlikely to hold in practice. LMM and MI give unbiased estimates under all four considered scenarios, provided that an interaction of intervention indicator and covariate is included in the model when the covariate effects are different in the two intervention groups. MI gives slightly overestimation of average standard error, which leads to a decrease in power. 
\end{abstract}
\section{Introduction}
In cluster randomised trials (CRTs), identifiable clusters of individuals such as villages, schools, medical practices -  rather than individuals - are randomly allocated to each of intervention and control groups, while individual-level outcomes of interest are observed within each cluster. The number of clusters and/or the cluster sizes in each intervention group might be different. CRTs with equal number of clusters in each intervention group with constant cluster size are known as balanced CRTs. One important characteristic of CRTs is that outcomes of individuals within the same cluster may exhibit more similarity compared to the outcomes of individuals in the other clusters, which is quantified by intraclass correlation coefficient (ICC), denoted by $ \rho $. In practice, the resulting value of ICC is typically ranges form 0.001 to 0.05 and it is rare to have ICC above 0.1 \cite{murray_blitstein2003}. Small values of ICC can lead to a substantial amount of variance inflation factors and cannot be ignored \cite{Donnerandklar2000,murray1998}. CRTs are being increasingly used in the fields of health promotion and  health service research. Reasons for such popularity may include the nature of intervention that itself may dictate its application at the cluster level, less risk of intervention contamination and high administrative convenience \cite{Donner1994}. It is well known that the power and precision of CRTs are lower relative to trials that individually randomise the same number of individuals. In spite of having this limitation, the advantages associated with CRTs are perceived by researchers to outweigh the resulting cost in statistical power and precision in some situations. 

Attrition is common in CRTs which leads to missing outcome data, that often create a problem in the analysis of such trials. Not only do they cause a loss of information and as a result usually reduce statistical power of the study, but also they might be a potential source of bias in the parameter estimates \cite{woodwhite2004}. Handling missing data in CRTs is complicated by the fact that data are clustered. Inadequate handling of the missing data may result in misleading inferences \cite{sterne2009}. A systematic review \cite{DiazOrdaz2014} revealed that, among all CRTs published in English in 2011, 72\% trials had missing values either in outcomes or in covariates or in both. Only 34\% CRTs of them reported how they handled missing data. One of the reasons may be that the methodological development for dealing with missing data in CRTs has been relatively slow in spite of the increasing popularity of CRTs. 

The impact of missing data on estimation and inference of a parameter of interest depends on the mechanism that caused missing data, the method used to handle missing data, and the choice of statistical methods used for data analysis. In CRTs, baseline covariates that might be related to the outcome of interest are often collected and these are sometimes incorporated into the analysis, known as covariate adjusted analysis. In  some CRTs, a plausible working assumption may be that missingness in outcomes depends on covariates measured at baseline, but not on the outcome itself, known as covariate dependent missingness (CDM).

This paper addresses the question of under which conditions, using complete case analysis (CCA), unadjusted cluster-level analysis, baseline covariate adjusted cluster-level analysis and linear mixed model (LMM) analysis are valid when there is missingness in a continuous outcome, with the probability of missingness depending on baseline covariate values. We also compare the performance of these methods using CCA with the performance of them applied to multiply imputed datasets. Baseline covariates in CRTs could be either individual-level or cluster level. In this paper we restrict attention to individual-level baseline covariates, and throughout this paper whenever we say baseline covariate, we mean baseline individual-level covariate, unless stated otherwise. 

This paper is organised as follows. Section \ref{analysis_of_CRTs_with_complete_data} presents a brief review of the approaches to the analysis of CRTs with complete data. In Section \ref{missingness_mechanism_for_CRTs}, the assumed missingness mechanism for CRTs is described alongside a discussion about the previous works on missing data in CRTs. Section \ref{methods_of_handling_missing_data_in_CRTs} describes the methods of handling missing data in CRTs. In Section \ref{analysis_of_CRTs_with_missing_data}, we give results which show under what conditions the various analyses give valid inferences. Section \ref{simulation_study} describes a simulation study and presents the results. In Section \ref{discussion_conclusion}, we conclude with some discussion.  
\section{Analysis of CRTs with complete data}
\label{analysis_of_CRTs_with_complete_data}
In this section, before considering the analysis of CRTs with missing outcome data,  we describe the two broad approaches to the analysis of CRTs in the absence of missing data. These are cluster-level analysis and individual-level analysis.
\subsection{Cluster-level analysis}
Cluster-level analysis can be done in two ways: unadjusted cluster-level analysis and baseline covariate adjusted cluster-level analysis. This approach can be explained as a two-stage process. In the first stage of unadjusted analysis,  a relevant summary measure of outcomes is calculated for each cluster. Then, in the second stage, the two sets of cluster specific summary measures obtained in the first stage are compared using  appropriate statistical methods. The most common one is the standard $ t-$test for two independent samples (here referred to as cluster-level $ t- $ test) with degrees of freedom (DF) equal to the total number of clusters in the study minus two. The basis of using this test is that the resulting summary measures are statistically independent, which is a consequence of the clusters being independent of each other. In the case of baseline covariate adjusted analysis, an individual-level regression analysis is carried out at the first stage of analysis without considering clustering of the data \cite{Gail1988, hayes2009}. In this first stage, all covariates, except the intervention indicator, are considered as explanatory variables into the regression model. Individual level residuals are then used to calculate the cluster-specific summary measures, which are then compared using cluster-level $ t- $test in the second stage of analysis to evaluate the intervention effect adjusted for baseline covariates. The main purposes of adjusting for covariates are to increase the credibility of the trial findings by demonstrating that any observed intervention effect is not attributed to the possible imbalance between the intervention groups in term of baseline covariates, and to improve the statistical power \cite{Hernandez2004}.  
\subsection{Individual-level analysis} 
In individual-level analysis, a regression model is fitted to the individual-level outcomes, with allowance for the fact that observations within the same cluster are correlated.  Two widely accepted approaches are random effects models and generalised estimating equations (GEE). Random effects models take into account of between-cluster variability using cluster-level effects which are assumed to follow a specified  probability distribution. The parameters of the regression model are estimated using maximum likelihood methods together with intervention effect and other covariates effects, if any. Depending on  the parameter of interest and the type of outcome, the most commonly used  random effects models are linear mixed model (LMM) for continuous outcomes, random effects Poisson regression model for event rates and random effects logistic regression (RELR) model for binary outcomes \cite{hayes2009}.  Generalised estimating equations (GEE) offers an alternative to random effects models that take into account the correlation among the outcomes of the same cluster using a working correlation matrix. GEEs are easier to fit than random effects models when the outcome is binary. The GEE method gives estimates of the marginal (also known as  population averaged) intervention effect, whereas RELR gives  estimates of the conditional (also known as cluster-specific) intervention effect.  In GEE, a reasonable large number of clusters in each group is needed to get reliable results and it is recommended to have at least 40 clusters in the study to get reliable standard errors of the estimates \cite{Murray2004}.  With small number of clusters in each group, GEE may give underestimated standard errors of the estimated intervention effect using the sandwich variance estimator, that may lead to elevated Type I error. 

The adjusted $ t- $test, proposed by Donner and Klar (2000) \cite{Donnerandklar2000}, is a alternative approach to test the intervention effect for quantitative outcomes by comparing the means of intervention groups using individual-level data. This test is a simple extension of the standard $ t- $test, which allows for correlation in outcomes. The adjusted $ t- $test and the cluster-level $ t- $test are identical for balanced CRTs.
\section{Missingness mechanism assumptions for CRTs}
\label{missingness_mechanism_for_CRTs}
In statistical analysis, if there are missing values, an assumption must be made about missingness mechanism, which refers to the relationship between missingness and the values of the variables in the data \cite{littlerubin2002}. According to Rubin's framework \cite{Rubin1976}, a missingness mechanism can be classified as (i) missing completely at random (MCAR), where the probability of a value being missing is independent of the observed and unobserved data, (ii) missing at random (MAR), where conditioning on the observed data,  the probability of a value being missing is independent of the unobserved data, and (iii) missing not at random (MNAR), where the probability of value being missing depends on both observed and unobserved data. In this paper, we will consider the common setting where some outcomes (continuous) are missing. 

In CRTs, an assumption that may sometimes be plausible is that missingness in outcomes depends on covariates measured at baseline, but conditional on these, not on the outcomes itself, known as covariate dependent missingness (CDM). For example, blood pressure outcome data could be CDM if missingness in blood pressure measurement depends on covariates (e.g. age, BMI or weight), but given these, not on the blood pressure measurement itself. CDM is an example of a MAR mechanism  when covariates are fully observed. 

Let $ Y_{ijl} $ be a continuous outcome of interest for the $ l $th $ (l=1,2,\ldots,m_{ij}) $ individual in the $ j $th $(j=1,2,\ldots,k_i) $ cluster of the intervention group $ i $ $(i=1,2) $, where $ i=1 $ corresponds to control group and $ i=2 $ corresponds to intervention group. We assume that the $ Y_{ijl} $ follow a linear mixed model given by
\begin{equation}
\label{model}
Y_{ijl}=\alpha_i+ \beta_i X_{ijl}+\delta_{ij}+\epsilon_{ijl},
\end{equation}
where $ \alpha_i $ is a constant for $ i $th intervention group, $ X_{ijl} $ is a baseline covariate value for $ (ijl) $th individual, $ \beta_i $ is the effect of covariate $ X $ on $ Y $ in intervention group $ i $, $ \delta_{ij} $ is the $ (ij) $th cluster effect and $ \epsilon_{ijl} $ is the individual error term. We also assume that the cluster effect $ (\delta_{ij}) $ and the individual error $ (\epsilon_{ijl}) $ are statistically independent; and  $ \text{E}\left( \delta_{ij}\right)=0 $, $ \text{Var}\left( \delta_{ij}\right)=\sigma^2_b $ and $ \text{E}\left( \epsilon_{ijl}\right)=0 $, $ \text{Var}\left( \epsilon_{ijl}\right)=\sigma^2_w $, where $ \sigma^2_b $ and $ \sigma^2_w $ are the between-cluster variance and within-cluster variance, respectively. Later we will sometimes make normality assumptions on these random effects/random errors. 
Suppose the covariate $ X $ has mean $ \mu_x $. Then
\begin{equation}\nonumber
\mbox{E}\left(\bar{Y}_{i} \right)= \alpha_i+\beta_i\mu_{x}=\mu_{i} (\text{say}),
\end{equation}
where $ \bar{Y}_i=(1/k_i) \sum_{j=1}^{k_i}(1/m_{ij})\sum_{l=1}^{m_{ij}}Y_{ijl}=(1/k_i) \sum_{j=1}^{k_i}\bar{Y}_{ij}$. Here,  $ \bar{Y}_i $ and $ \bar{Y}_{ij} $ are the mean outcome of the $ i $th intervention group and the $ (ij) $th cluster, respectively. With complete data, the cluster-level analysis estimate of the intervention effect, say $ \hat{\theta} $, is then calculated as
\begin{equation}\nonumber
\hat{\theta}=\bar{Y}_1-\bar{Y}_2.
\end{equation}
With complete data, this estimator is unbiased for the true intervention effect, that is,  
\begin{equation}\nonumber
\text{E}(\hat{\theta})=\mu_1-\mu_2. 
\end{equation}  
Suppose there are some missing values for outcome $ Y $. Define a missing data indicator $ R_{ijl} $ such that
\begin{equation*}
R_{ijl}=\begin{cases}
1, \text{~if~} Y_{ijl} \text{~is observed~}\\
0, \text{~if~} Y_{ijl} \text{~is missing~}.
\end{cases}
\end{equation*}
Then $ \sum_{l=1}^{m_{ij}}R_{ijl} $ is the number of observed values in the $ (ij) $th cluster. The CDM assumption can then be expressed as 
\begin{equation}
 P(R_{ijl}=0|\boldsymbol{Y}_{ij}, \boldsymbol{X}_{ij}) =P(R_{ijl}=0|X_{ijl}),\nonumber
\end{equation}
where $ \boldsymbol{Y}_{ij}=(Y_{ij1},Y_{ij2}, \ldots, Y_{ijm_{ij}}) $ and $ \boldsymbol{X}_{ij}=(X_{ij1},X_{ij2}, \ldots, X_{ijm_{ij}}) $ are the  vectors of the outcomes and the covariate values, respectively, in the $ (ij) $th cluster. In other words, the missingness of the $ (ijl) $th individual's outcome $Y_{ijl} $ depends only on that individual's covariate value $ X_{ijl} $.  
\section{Methods of handling missing data in CRTs}
\label{methods_of_handling_missing_data_in_CRTs}
Common approaches for handling missing data in CRTs include complete case analysis (CCA), single imputation and multiple imputation (MI). In this  paper, we focused on CCA and MI since they are the most commonly used methods now for handling missing data. Next we describe these two approaches briefly considering missingness in outcomes only.
\subsection{Complete case analysis}
In complete case analysis (CCA), only cases with outcome observed are considered in the analysis, while cases with missing outcome are excluded. It is widely used because of its simplicity and is usually the default method of most statistical packages. It is well known that CCA is valid if data are MCAR  or if missingness is independent of the outcome, conditional on covariates \cite{littlerubin2002}. Likelihood based CCA is valid under MAR, if missingness is only in the outcome and all predictors of missingness are conditioned on in the model \cite{littlerubin2002}. CCA is also valid under MNAR mechanisms where missingness in a covariate is dependent on the value of that covariate, but is conditionally independent of outcome \cite{White2010, Bartlett2014_biostat} 
\subsection{Single imputation}
\label{single.imputation}
Instead of discarding incomplete cases, single imputation imputes a single value for each missing outcome and creates a complete data set. Two choices for single imputation that have been considered in CRTs are group mean imputation and cluster mean imputation \cite{taljaard2008}. In the first case, missing outcomes in each intervention group are replaced by the mean outcome calculated using complete cases (CCs) pooled across clusters of that group. This approach reduces the variability among the clusters means \cite{taljaard2008} and, therefore, gives inflated Type I error. In cluster mean imputation, missing outcomes in each cluster are replaced by the mean outcome calculated using CCs of that cluster.  This approach has been suggested as a good approach for handling missing outcomes under MCAR by  Taljaard \textit{et al.} \cite{taljaard2008}. They showed that cluster mean imputation gives Type I error close to nominal level (5\%) under MCAR, using adjusted $ t-$test with balanced CRTs. However, under MAR or CDM, adjusted $ t-$test may not be valid.  We note that, with balanced CRTs, the cluster-level $ t- $test and the adjusted $ t-$test are identical with cluster mean imputation  since after imputation the cluster sizes become constant and the cluster means remain unchanged.
Consequently, our later results for the validity of cluster level t-test can also be applied to infer the validity of results after using cluster mean imputation. One problem with cluster mean imputation is that it distorts the estimates of between-cluster variability  and within-cluster variability, which often are of interest.
\subsection{Multiple imputation }
 Multiple imputation (MI), first proposed by Rubin (1987) \cite{rubin1987}, is a method of filling in the missing outcomes multiple times by simulating from an appropriate model. The aim of imputing multiple times is to allow for the uncertainty about the missing outcomes due to the fact that the imputed values are sampled draws for the missing outcomes. A sequence of $ Q $ imputed data sets are obtained by replacing each missing outcome by a set of $ Q\ge 2 $ imputed values that are simulated from an appropriate distribution or model. Each of the $ Q $ data sets are then analysed as a completed data set using a standard method. The results from the $ Q $ imputed data sets are then combined using Rubin's rules \cite{rubin1987}. The combined inference is based on a $ t- $distribution with DF given by
\begin{equation}
\nu=(Q-1)\left( 1+\frac{Q}{Q+1} \frac{W_{\text{MI}}}{B_{\text{MI}}}\right)^2,
\label{df_com}
\end{equation}
where $ B_{\text{MI}} $ is the between-imputation variance and $ W_{\text{MI}} $ is the average  within-imputation variance. This formula for DF is derived under the assumption that the complete data DF, $ \nu_\text{com} $, is infinite \cite{Barnard1999}. 

In CRTs, $ \nu_\text{com} $ is usually small  as it is based on the number of clusters in each intervention group rather than the number of individuals. For unadjusted cluster-level analysis and individual-level baseline covariate adjusted cluster-level analysis, $ \nu_\text{com} $  is calculated as $ k_1+k_2-2 $ for statistical inference using cluster-level $ t- $test \cite{hayes2009} and adjusted $ t- $test \cite{taljaard2008}. An adjustment is made to the $ \nu_\text{com} $ to adjust for cluster-level baseline covariates using cluster-level analysis. In this case, we reduce the complete data DF from $ \nu_\text{com}= k_1+k_2-2 $ to $ \nu_\text{com}= k_1+k_2-2-p $, where $ p $ is the number of parameters corresponding to the cluster-level covariates in the first stage regression model   \cite{hayes2009}.

 When $ \nu_{\text{com}} $ is small and there is a modest proportion of missing data,  the repeated-imputation DF,  $ \nu $ (given in \ref{df_com}), for reference $ t- $ distribution can be much higher than $ \nu_{\text{com}} $, which is not appropriate \cite{Barnard1999}. In such a situation, a more appropriate DF, $ \nu_{\text{adj}} $, proposed by Barnard and Rubin (1999) \cite{Barnard1999}, is calculated as
\begin{equation}
\nu_{\text{adj}}=\left(\frac{1}{\nu} +\frac{1}{\hat{\nu}_{\text{obs}}}\right)^{-1} \le  \nu_{\text{com}},
\label{df.adj} 
\end{equation}
where
\begin{equation} \nonumber
\hat{\nu}_{\text{obs}}= \left( 1+\frac{Q+1}{Q} \frac{B_{\text{MI}}}{W_{\text{MI}}} \right)^{-1} \left( \frac{\nu_{\text{com}}+1}{\nu_{\text{com}}+3}\right)\nu_{\text{com}}.
\end{equation}
At least four different types of MI have been used in CRTs \cite{DiazOrdaz2014}. These are \textit{standard }MI which ignores clustering, \textit{fixed effects } MI  which includes a fixed effect for each cluster in the imputation model, \textit{random effects} MI where clustering is taken into account through random effects in the imputation model and \textit{within-cluster} MI where standard MI is applied within each cluster. Andridge   \cite{Andridge2011} showed, with balanced CRTs under MCAR and MAR missingness in a continuous outcome with a single covariate in addition to intervention indicator, that MI models that incorporate clustering using fixed effects for cluster can result in a serious overestimation of variance of group means and this overestimation is more serious for small cluster sizes and small ICCs. This overestimation of variance results in a decrease in power, which is particularly dangerous for CRTs which are often underpowered \cite{Andridge2011}.  The MI using random effects for cluster  gave slight overestimation of variance of group means for very small values of $ \rho $. Andridge also showed that using an MI model that ignores clustering can lead to severe  underestimation of the MI variance for large values of $ \rho  $ ($ > $0.005). This underestimation of variance leads to inflated Type I error. 

Taljaard \textit{et al.} \cite{taljaard2008} examined the performance of MI in a simple set-up considering balanced CRTs where there are no covariates except intervention indicator using standard regression imputation, which ignores clustering, and random effects MI  which does account for intraclass correlation.  They also considered the Approximate Bayesian Bootstrap (ABB) procedure, proposed by Rubin and Schenker \cite{RubinSchenker1986}, as a non-parametric MI. In ABB, sampling from the posterior predictive distribution of missing data is approximated by first generating a set of plausible contributors drawn with replacement from the observed data, and then imputed values are drawn with replacement from the possible contributors.  Two possible uses of ABB in CRTs are pooled ABB and within-cluster ABB, where the set of possible contributors are sampled from all observed values across the clusters in each group or  from observed values in the same cluster, respectively. They showed that none of these four MI procedures tend to yield better power compared to the power of adjusted $ t-$test using no imputation and cluster mean imputation under MCAR.

In the case of missing outcome under MAR for non-clustered trials, Groenwold \textit{et al.} \cite{Groenwold2012} showed that CCA with covariate adjustment and MI give similar estimates so long as the same set of predictors of missingness are used. The CCA with covariate adjustment has the advantage of being easier to apply compared to MI. 

Ma \textit{et al.} \cite{Ma2011} examined the performance of within-cluster MI and fixed effect MI to estimate the intervention effect and its confidence interval through a simulation study in the case of missing binary outcomes in CRTs. In within-cluster MI,  standard MI strategies, which includes logistic regression method, propensity score method and normal based Markov Chain Monte Carlo method, were applied within each cluster. The fixed effect MI considers a fixed effect for each cluster.  They considered  MCAR and CDM mechanisms in balanced CRTs. They showed that all these strategies give quite similar results for low percentages of missing data or for small value of ICC. But with high percentage of missing data, the standard MI strategies underestimate the variance of the intervention effect which may result in inflated Type I error.  In a subsequent study again by Ma \textit{et al.} \cite{Ma2013}, the performance of GEE and RELR were compared considering a balanced CRT with missing binary outcomes using standard MI and within-cluster MI. GEE was found to perform well -  using complete case analysis with small proportion of missing data; using standard MI with variance inflation factor (VIF) less than 3; and using within-cluster MI  with VIF $ \ge 3$ and cluster size at least 50. They also concluded that RELR performs well only when the percentage of missing data is small and it doesn't perform well with standard MI or with within-cluster MI. In another study, Caille \textit{et al.} \cite{Caille2014} demonstrated different MI strategies for handing missing binary data in CRTs to assess bias, standard error and coverage probability of the population averaged intervention effect. They considered CCA, baseline adjusted CCA, single and multiple imputation approaches. Results showed that MI with RELR model or with standard logistic regression model gave unbiased estimates of the population averaged intervention effect and good coverage probability as well, although the latter method produced slightly lower estimate of coverage probability. All of these studies done by Ma \textit{et al.} \cite{Ma2011,Ma2013} and Caille \textit{et al.} \cite{Caille2014} considered a relatively high  number of clusters in each intervention group.
\section{Analysis of CRTs with missing data}
\label{analysis_of_CRTs_with_missing_data} 
In this section, we describe the unadjusted cluster-level analysis, baseline covariate adjusted cluster-level analysis and linear mixed model analysis methods using CCs, and derive conditions under which they give valid inferences under CDM assumption. 
 \subsection{Unadjusted cluster-level analysis using CCs}
 \label{unadj_ccs}
The mean of the observed outcomes in the $ i $th intervention group can be calculated as
\begin{equation}
\bar{Y}_{i}^{\text{obs}} = \frac{1}{k_i}\sum_{j=1}^{k_i}\bar{Y}_{ij}^{\text{obs}},\nonumber
\end{equation}
where $\bar{Y}_{ij}^{\text{obs}}= \left( {1}/{\sum_{l=1}^{m_{ij}}R_{ijl}}\right) \sum_{l=1}^{m_{ij}} R_{ijl}Y_{ijl}$ is the observed mean of $ (ij) $th cluster.  The estimate of intervention effect is given by
\begin{equation}
\label{estuadj}
\hat{\theta}^{\text{obs}}=\bar{Y}_1^{\text{obs}}-\bar{Y}_2^{\text{obs}}.
\end{equation}
In Appendix \ref{appenA}, we show that
\begin{equation}
\mbox{E}\left( \hat{\theta}^{\text{obs}} \right) =  \mu_1-\mu_2 +\beta_1\left( \mu_{x11}-\mu_{x}\right) -\beta_2\left( \mu_{x21}-\mu_{x}\right),
\label{exp:unadj}
\end{equation}
and
\begin{equation}
\text{Var}\left(  \hat{\theta}^{\text{obs}} \right) 
= \sum_{i=1}^{2} \frac{1}{k_i}\left( \beta_i^{2}\sigma^2_{\bar{x}_{i1}}+\sigma_{b}^{2}+\frac{\sigma_{w}^{2}}{\eta_i}\right),
\end{equation}
where $ \mu_{xi1} $ is true mean of the baseline covariate $ X $ in the $ i $th intervention group among those individuals with observed outcomes, $ \sigma^2_{\bar{x}_{i1}} $ is the variance of $ \bar{x}_{i1} $, the sample mean of $ X $ values in the $ i $th intervention group among those individuals with observed outcomes, and $ {1}/{\eta_i}=\text{E}\left(  {1}/{\sum_{l}^{}R_{ijl}} \right)$. Hence, the estimator (\ref{estuadj}) will be unbiased if $ \beta_1=\beta_2 $ and $ \mu_{x11}=\mu_{x21} $. In other words, the unadjusted cluster-level analysis is unbiased only if the two intervention groups have the same missingness mechanisms and the same covariate effects. 
\subsection{Adjusted cluster-level analysis using CCs}
\label{adj_ccs}
 Let $ \hat{\epsilon}_{ijl} $ be the estimated residual for $ (ijl) $th individual. Then 
\begin{equation}
\hat{\epsilon}_{ijl}=Y_{ijl}-\hat{Y}_{ijl}\nonumber,
\end{equation}
where $ \hat{Y}_{ijl}=\gamma+\lambda X_{ijl} $ is the predicted outcome for the $ (ijl) $th individual.  
The mean of the observed residuals of the $ i $th group is given by
\begin{equation}
{\bar{\hat{\epsilon}}_{i}}^\text{obs}= \frac{1}{k_i} \sum_{j=1}^{k_i} {\bar{\hat{\epsilon}}_{ij}}^\text{obs}\nonumber,
\end{equation}
where ${\bar{\hat{\epsilon}}_{ij}}^\text{obs}={1}/\left( {\sum_{l=1}^{m_{ij}}R_{ijl}}\right) \sum_{l=1}^{m_{ij}}R_{ijl} \hat{\epsilon}_{ijl} $ is the mean of observed residuals of the $ (ij) $th cluster. The  baseline covariate adjusted estimator of  intervention effect is given by
\begin{equation}
\label{esti_adj}
\hat{\theta}_{\text{adj}}^{\text{obs}} = {\bar{\hat{\epsilon}}_{1}}^\text{obs} - {\bar{\hat{\epsilon}}_{2}}^\text{obs}.
\end{equation}
We show in Appendix \ref{AppenB} that
\begin{equation}
\mbox{E}\left( \hat{\theta}_{\text{adj}}^{\text{obs}} \right) 
= \mu_1-\mu_2 +\beta_1\left( \mu_{x11}-\mu_{x}\right) -\beta_2\left( \mu_{x21}-\mu_{x}\right)+\lambda\left( \mu_{x21}-\mu_{x11} \right).
\label{exp:adj}
\end{equation}
Hence, the estimator (\ref{esti_adj}) will be unbiased if (i) $ \beta_1=\beta_2  $ and $ \mu_{x11}=\mu_{x21} $, or if (ii) $ \lambda=\beta_1=\beta_2  $. Equation (\ref{exp:adj} ) is derived (see Appendix \ref{AppenB}) assuming fixed values of $ \gamma \text{~and~}\lambda $ instead of their estimates. In practice, $ \gamma $ and $ \lambda $ are unknown and must be estimated by fitting a first stage regression model for the observed outcomes, where all covariates except the intervention indicator are included in the regression model, and clustering is ignored. We are not worried about the estimate of the intercept parameter $ \gamma $ since the expression (\ref{exp:adj}) is independent of $ \gamma $. If $ \lambda $ is estimated consistently, then $ \hat{\theta}_{\text{adj}}^{\text{obs}}  $ will be a consistent estimator of intervention effect when in truth $\lambda=\beta_{1}=\beta_{2}$. The estimator of $ \lambda $, say $ \hat{\lambda} $, is calculated using CCs, and will be unbiased (and therefore consistent)  if $ R_{ijl}\perp Y_{ijl}|X_{ijl} $. This is true only when the two intervention groups have the same missingness mechanisms and the same covariate effects. Therefore, assuming CDM, the baseline covariate adjusted cluster-level analysis is consistent only if the two intervention groups have the same covariate effects and the same missingness mechanisms.

The variance of the estimator (\ref{esti_adj}) can be written as (see Appendix \ref{AppenB} for derivation)
\begin{equation}
\mbox{Var}\left(\hat{\theta}_{\text{adj}}^{\text{obs}}  \right)
= \sum_{i=1}^{2} \frac{1}{k_i}\left( \left( \beta_i-\lambda  \right)^2\sigma^2_{\bar{x}_{i1}} +\sigma_b^2+\frac{\sigma_w^2}{\eta_i}  \right).
\label{var:adj}
\end{equation}
This shows that when $ \beta_1=\beta_2 $ and the missingness mechanisms are the same in the two intervention groups, in order for the estimator
$( \ref{esti_adj}) $ to have minimum variance one should replace the unknown $ \lambda $ by an estimate
of $ \beta_1=\beta_2 =\beta (\text{say})$.
\subsection{Linear mixed model using CCs}
Let $ Z $ be the intervention indicator which is zero for control group and is one for intervention group. When it is assumed that the two intervention groups have the same covariate effects, we fit a LMM with fixed effects of $ X $ and $ Z $, and a random effect for cluster. Then the estimate of the coefficient of $ Z $ will be the estimated intervention effect accounting for $ X $. 

If one thinks that the covariate effects could be different in the two intervention groups, an interaction of $ X $  and $ Z $ must be included in the model. This implies that the intervention effect varies with $ X $. Then the estimate of the intervention effect at the mean value of $ X $ is known as average intervention effect. Let $ X^* $ denote the empirically centred variable $ X-\bar{X} $. If the covariate effects are assumed to be different in the two groups, we fit a LMM with fixed effects of $ X^* $, $ Z $ and their interaction, and a random effect for cluster. The estimate of the coefficient of $ Z $ will then be the estimated average intervention effect.    

In the general theory of LMM, the variance of the fixed effects parameter estimates, which are calculated based on their  asymptotic distributions, are known to be underestimated for small sample sizes \cite{kenward1997}. In practice, for testing hypotheses about fixed-effects parameters, this resulting downward bias is often handled by using approximate $ t- $statistic and $ F- $statistic \cite{Verbeke2000}. An approximate $ t-$test can be obtained by approximating the distribution of $ t- $statistic by an appropriate $ t- $distribution.  Satterthwaite \cite{Satterthwaite1941} proposed an approximation to calculate the DF of the $ t- $distribution. For testing hypotheses of the form $ H_0: \mathbf{L}\boldsymbol{\kappa}=\mathbf{0} $, where $ \boldsymbol{\kappa} $ is a vector of fixed-effects parameters and $ \mathbf{L} $ is any known matrix, Kenward and Roger \cite{kenward1997} suggested a scaled Wald statistic as well as an $ F $ approximation of its sampling distribution that performs well for small sample size. The suggested statistic uses an adjusted estimate of the variance-covariance matrix that has reduced small sample bias than the standard asymptotic variance-covariance estimate. The numerator DF of the approximate $ F- $distribution equals rank$ (\mathbf{L}) $ and the denominator DF is calculated via a satterthwaite - type approximation \cite{Verbeke2000}. For only one fixed effect in the model, Kenward and Roger's approximation essentially recovers Satterthwaite's approximation \cite{kenward1997}. Both these approximation are applicable for linear mixed models and related multivariate normally based models \cite{Faes2009}. As far as we are aware no study has been done to compare these two approximations in CRTs. We have conducted a simulation study to examine the performance of these two approximations for DF and the complete data DF, $ \nu_\text{com}=k_1+k_2-2  $, for testing hypothesis of fixed effect in the case of balanced CRTs with only covariate $ Z $ (not $ X $). Missingness in outcome was considered under MCAR and CDM (missingness depends on intervention indicator).  A LMM was fitted using both complete data and CCA. We used $ z-$test and Wald $t-  $test (using Satterthwaite, Kenward-Roger's and  $ \nu_\text{com} $  DF) for testing hypothesis about the fixed effect of $ Z $. Results (not presented in the paper) showed that the $ z- $test tends to have inflated Type I error rates for small $ k $ using both complete data and CCA. The Wald $ t-$test with all the three DFs yields Type I error rates close to the nominal level at all considered values of $ k,m $ and $ \rho $.  Both tests result in acceptable Type I error rate for $ k>15 $.
\section{Simulation Study}
\label{simulation_study}
A simulation study was conducted to investigate the performance of unadjusted and baseline covariate adjusted cluster-level analyses, LMM and MI under baseline covariate dependent missingness in outcomes. The average estimate of intervention effect, its average estimated standard error (SE) and  coverage probability were calculated and compared to each other. We considered balanced CRTs, where the two intervention groups have  equal number of clusters $ (k_i=k) $ and constant cluster size $ (m_{ij}=m) $.
\subsection{Data generation and analysis}
For each individual in the study a single covariate value $ X $ was generated independently as $ X\sim N(0,1) $. Since $ \sigma_x^2=1 $, we can write the coefficient of $ X $ in (\ref{model})  as $ \beta_i=\tau_i\sigma_y $, where $ \sigma_y^2 $ is the total variance of $ Y $ within each intervention group and $ \tau_i $ is the correlation coefficient between $ Y $ and  $ X $ in intervention group $ i $. We  fixed $ \sigma_y^2=100 $, $ \alpha_1=20 $ and $ \alpha_2=25 $. Then the outcome $ Y $ was generated using the model 
\begin{equation}
Y_{ijl}=\alpha_i + \tau_i\sigma_y X_{ijl}+\delta_{ij}+\epsilon_{ijl}\nonumber,
\end{equation}
where $ \delta_{ij}\sim  N(0,\rho \sigma_y^2) $ and $ \epsilon_{ijl}\sim N(0, (1-\tau_i^2-\rho)\sigma_y^2) $.  We chose the cluster size $ m=30 $ for each cluster. Parameters that were varied in generating the data include the number of clusters in each group, $ k=(5,10,20,30) $ and the unconditional ICC, $\rho=(0.001,0.05,0.1) $. In addition, we considered $ \tau_1=\tau_2=0.5 $ to have the same covariate effects in the two intervention groups, and $ \tau_1=0.4 $, $ \tau_2=0.6 $ to have  different covariate effects in the two  intervention groups.  The missing data indicators $ R_{ijl} $  under CDM assumption were generated, independently for each individual, according to a logistic regression model
\begin{equation}
\text{logit}\left( R_{ijl}=0\big | \boldsymbol{Y}_{ij},\boldsymbol{X}_{ij}\right) =\phi_{i0}+\phi_{i1} X_{ijl}.
\label{miss_equ}
\end{equation}
The intercept $ \phi_{i0} $ and slope $ \phi_{i1} $ were chosen so that $ \text{E}_{jl}\left( R_{ijl}\right)=p_i  $, where $ p_i $ is the desired proportion of missing values in intervention group $ i $. The degree of correlation between missingness and covariate depends on the value of $ \phi_{i1}$. We used $ \phi_{11}=\phi_{21} =1$, which gives odds ratio=$ \exp(1)=2.72 $, that is, the odds ratio for having a missing outcome $ (Y) $ is  2.72 associated with a one unit increase in the covariate $ (X) $ value.  Missing data indicators were then imposed to each generated complete data to get the incomplete data.

Four possible scenarios were considered depending on the effects of covariate on outcome and missingness mechanism in the two intervention groups.
\begin{enumerate}
	\item The two intervention groups have the same missingness mechanisms and the same covariate effects.
	\item The two intervention groups have different missingness mechanisms but the same covariate effects.
	\item The two intervention groups have different covariate effects but the same missingness mechanisms.
	\item The two intervention groups have different missingness mechanisms and different covariate effects.
	\end{enumerate}
In the first and third scenarios, we chose $ \phi_{10}=\phi_{20} =-1$ and $ \phi_{11}=\phi_{21} =1$ so that there was 30\% missing outcomes in both groups. In the second and fourth scenarios, we chose $ \phi_{10}= -1,  \phi_{11} =1$ so that there was 30\% missing outcomes in the control group, and $ \phi_{20}= 0.5,  \phi_{21} =1$ so that there was  60\% missing outcomes in the intervention group. Each generated incomplete data set was then analysed using unadjusted cluster-level analysis, baseline covariate adjusted cluster-level analysis and LMM using CCs. We  included the interaction between intervention indicator  and covariate into the LMM in the third and fourth scenarios, where the two intervention groups have different covariate effects.  

The R package \texttt{jomo} \cite{matteo} was used to multiply impute each generated incomplete data set using MI with number of imputations 20. A random intercept LMM was used as the imputation model so that the imputation model was correctly specified. We used 200 burn-in iterations and 10 iterations between two successive draws after examining, respectively,  the convergence of the posterior distributions of the parameters estimates of the imputation model and the plots of their  autocorrelation functions.   The completed data sets were then analysed using LMM. An interaction between intervention indicator and covariate was included in both the imputation model and the analysis model when the covariate effects were different in the two intervention groups. We always used the restricted maximum likelihood estimation method to fit the LMM. The  Wald $ t- $test with adjusted DF, given in equation \ref{df.adj}, with $ \nu_\text{com}=2(k-1) $ was used to test the null hypothesis of intervention effect. We had some convergence warning messages ( 33-50 out of 10000 simulations) when the LMM was fitted using the R package \texttt{lme4} \cite{lme4}.
\subsection{Results}
Empirical average estimates of intervention effect, average estimated standard errors (SEs) and coverage probabilities of nominal 95\% interval over 10000 simulation runs for all the four scenarios  are presented in Tables \ref{tab:case1} to \ref{tab:case4}. 

When the two intervention groups have the same missingness mechanisms and the same covariates effects, both the unadjusted and baseline covariate adjusted analyses gave
\begin{table}[H]
	\caption{Simulation results-the two intervention groups have the same missingness mechanisms and the same covariate effects. Empirical average estimates of intervention effect, average estimated SEs and coverage probabilities of nominal 95\% interval over 10000 simulation runs for unadjusted cluster-level analysis (Uadj), baseline covariate adjusted cluster-level analysis (Adj), linear mixed model (LMM), using CCA, and multiple imputation (MI). Monte-Carlo errors for average estimates and average estimated SEs are all less than 0.023 and 0.016, respectively. The true value of the intervention effect is 5.}
	\centering 
	\scalebox{0.85}{
	\begin{tabular}{c|c|cccc|cccc|cccc}
		\toprule
		\multirow{2}{*}{$ \rho $} &	\multirow{2}{*}{$ k $}	& \multicolumn{4}{c|}{Average Estimate}  & \multicolumn{4}{c|}{Average estimated SE} & \multicolumn{4}{c}{Coverage (\%)}\\
		\cline{3-14}
		&	& Uadj & Adj  & LMM & MI & Uadj & Adj &  LMM & MI & Uadj & Adj &  LMM & MI \\
		\midrule
		\multirow{4}{*}{0.1} 
		& 5  & 4.98 & 4.99 & 4.99 & 4.98 & 2.31 & 2.21 & 2.23 & 2.19 & 95.2 & 95.1 & 95.2 & 96.3 \\
		& 10 & 5.01 & 4.98 & 5.00 & 4.99 & 1.66 & 1.59 & 1.60 & 1.59 & 95.1 & 95.3 & 95.3 & 95.5 \\
		& 20 & 4.99 & 4.99 & 4.99 & 4.99 & 1.18 & 1.14 & 1.14 & 1.14 & 94.9 & 95.0 & 94.9 & 94.8 \\
		& 30 & 5.01 & 5.00 & 5.01 & 5.01 & 0.97 & 0.93 & 0.93 & 0.93 & 95.0 & 95.0 & 94.9 & 95.0 \\
		\hline
		\multirow{4}{*}{0.05}
		& 5  & 5.00 & 4.98 & 5.00 & 5.00 & 1.88 & 1.76 & 1.78 & 1.76 & 95.2 & 95.1 & 95.6 & 96.2\\
		& 10 & 5.01 & 5.00 & 5.01 & 5.01 & 1.35 & 1.28 & 1.28 & 1.26 & 95.1 & 95.2 & 95.1 & 95.4\\
		& 20 & 5.01 & 5.00 & 5.01 & 5.01 & 0.96 & 0.91 & 0.91 & 0.90 & 95.0 & 95.0 & 95.1 & 95.0\\
		& 30 & 4.99 & 4.99 & 4.99 & 4.99 & 0.79 & 0.75 & 0.74 & 0.74 & 95.0 & 95.0 & 95.0 & 95.0\\
		\hline
		\multirow{4}{*}{0.001}
		& 5  & 4.98 & 4.98 & 4.99 & 4.99 & 1.34 & 1.18 & 1.31 & 1.35 & 95.2 & 95.1 & 96.2 & 99.6\\ 
		& 10 & 5.01 & 5.00 & 5.01 & 5.01 & 0.96 & 0.85 & 0.90 & 0.93 & 95.1 & 95.1 & 96.8 & 97.8\\ 
		& 20 & 4.99 & 4.99 & 5.00 & 5.00 & 0.69 & 0.61 & 0.63 & 0.64 & 94.8 & 94.9 & 96.2 & 96.7\\ 
		& 30 & 5.00 & 5.00 & 5.00 & 5.00 & 0.56 & 0.50 & 0.51 & 0.52 & 95.1 & 95.3 & 96.2 & 96.8\\ 
		\bottomrule
	\end{tabular}}
	\label{tab:case1}
\end{table}
\noindent
 unbiased estimates of intervention effect  with coverage probabilities very close to the nominal level (see Table \ref{tab:case1}). But these two methods gave biased estimates of intervention effect if the two intervention groups have either different missingness mechanisms or different covariate effects or both (see Table \ref{tab:case2}, Table \ref{tab:case3} and Table \ref{tab:case4}, respectively). These results support our derived results in Section \ref{unadj_ccs} and Section \ref{adj_ccs}. These results also imply  
\begin{table}[H]
\caption{Simulation results-the two intervention groups have the different missingness mechanisms but the same covariate effects. Empirical average estimates of intervention effect, average estimated SEs and coverage probabilities of nominal 95\% interval over 10000 simulation runs for unadjusted cluster-level analysis (Uadj), baseline covariate adjusted cluster-level analysis (Adj), linear mixed model (LMM), using CCA, and multiple imputation (MI). Monte-Carlo errors for average estimates and average estimated SEs are all less than 0.025 and 0.017, respectively. The true value of the intervention effect is 5.} 
\centering
\scalebox{0.85}{
\begin{tabular}{c|c|cccc|cccc|cccc}
		\toprule
		\multirow{2}{*}{$ \rho $} &	\multirow{2}{*}{$ k $}	& \multicolumn{4}{c|}{Average Estimate}  & \multicolumn{4}{c|}{Average estimated SE} & \multicolumn{4}{c}{Coverage (\%)}\\
		\cline{3-14}
		&	& Uadj & Adj  & LMM & MI & Uadj & Adj &  LMM & MI & Uadj & Adj &  LMM & MI \\
		\midrule
		\multirow{4}{*}{0.1} 
		& 5  & 3.83 & 4.94 & 5.01 & 5.01 & 2.44 & 2.32 & 2.34 & 2.28 & 93.2  & 95.1 & 95.2  & 97.0\\
		& 10 & 3.81 & 4.94 & 5.03 & 5.03 & 1.76 & 1.67 & 1.68 & 1.66 & 89.9  & 95.4 & 95.2  & 95.5\\
		& 20 & 3.78 & 4.91 & 5.00 & 4.99 & 1.25 & 1.19 & 1.19 & 1.19 & 84.2  & 94.9 & 94.8  & 94.8\\
		& 30 & 3.79 & 4.93 & 5.01 & 5.01 & 1.02 & 0.98 & 0.98 & 0.98 & 79.1  & 95.4 & 95.3  & 95.4\\
		\hline
		\multirow{4}{*}{0.05} 
		& 5  & 3.77 & 4.90 & 4.98 & 4.98 & 2.04 & 1.90 & 1.94 & 1.92 & 91.7 & 94.9 & 95.7 & 98.3\\
		& 10 & 3.78 & 4.90 & 5.00 & 4.99 & 1.48 & 1.38 & 1.38 & 1.36 & 87.5 & 95.0 & 95.0 & 95.8\\
		& 20 & 3.76 & 4.92 & 4.98 & 4.98 & 1.05 & 0.98 & 0.98 & 0.97 & 79.4 & 95.2 & 95.1 & 95.1\\
		& 30 & 3.77 & 4.92 & 4.99 & 4.99 & 0.86 & 0.80 & 0.80 & 0.80 & 70.7 & 94.8 & 94.6 & 94.7\\
		\hline
		\multirow{4}{*}{0.001} 
		& 5  & 3.77 & 4.89 & 5.00 & 5.00 & 1.58 & 1.39 & 1.54 & 1.60 & 89.4 & 95.1 & 98.3 & 99.7 \\ 
		& 10 & 3.76 & 4.89 & 4.99 & 4.98 & 1.14 & 1.01 & 1.06 & 1.10 & 82.1 & 95.0 & 97.3 & 98.5 \\ 
		& 20 & 3.78 & 4.91 & 5.00 & 5.00 & 0.81 & 0.72 & 0.74 & 0.76 & 68.8 & 95.2 & 96.4 & 97.3 \\ 
		& 30 & 3.78 & 4.92 & 5.00 & 5.00 & 0.66 & 0.59 & 0.60 & 0.61 & 56.1 & 94.9 & 95.8 & 96.5 \\ 
		\bottomrule 
\end{tabular}}
\label{tab:case2}
\end{table}
\noindent
that the adjusted $ t- $test with cluster mean imputation (described in Section \ref{single.imputation} ) is not valid under CDM assumption unless the two intervention groups have the same missingness mechanisms and the same covariate effects. The bias in average intervention effect estimates could be in either direction. But, in this paper, we always have downward bias in the reported  intervention effect estimates. This is because we considered, in the data generation process, a positive correlation between covariate and outcome, and a positive association between covariate and probability of missingness in outcomes. As a result, a large value of outcome has higher chance of being missing compared to a low value of outcome. In our simulations the degree of bias was high if the two intervention groups have different covariate effects and it goes up if, in addition,  the two intervention groups have different missingness mechanisms (see Table \ref{tab:case3} and Table \ref{tab:case4}). LMM and MI gave unbiased estimates of intervention effect under all the four considered scenarios, provided that an interaction of intervention indicator and covariate was included in the model \hskip 0.3cm to 
\begin{table}[H]
	\caption{Simulation results-the two intervention groups have different covariate effects but the same missingness mechanisms. Empirical  average estimates of intervention effect, average estimated SEs and coverage probabilities of nominal 95\% interval over 10000 simulation runs for unadjusted cluster-level analysis (Uadj), baseline covariate adjusted cluster-level analysis (Adj), linear mixed model (LMM), using CCA, and multiple imputation (MI). Monte-Carlo errors for average estimates and average estimated SEs are all less than 0.024 and 0.016, respectively. The true value of the intervention effect is 5} 
	\centering
	\scalebox{0.85}{
	\begin{tabular}{c|c|cccc|cccc|cccc}
		\toprule
		\multirow{2}{*}{$ \rho $} &	\multirow{2}{*}{$ k $}	& \multicolumn{4}{c|}{Average Estimate}  & \multicolumn{4}{c|}{Average estimated SE} & \multicolumn{4}{c}{Coverage (\%)}\\
		\cline{3-14}
		&	& Uadj & Adj  & LMM & MI & Uadj & Adj &  LMM & MI & Uadj & Adj &  LMM & MI \\
		\midrule
		\multirow{4}{*}{0.1} 
		& 5  & 4.46 & 4.44 & 4.97 & 4.97 & 2.31 & 2.22 & 2.25 & 2.22 &  94.3 & 94.3 & 95.0 & 96.4\\
		& 10 & 4.50 & 4.49 & 5.01 & 5.02 & 1.66 & 1.59 & 1.61 & 1.60 &  93.7 & 93.6 & 94.7 & 94.8\\
		& 20 & 4.48 & 4.48 & 5.00 & 5.00 & 1.19 & 1.14 & 1.15 & 1.15 &  92.5 & 92.6 & 94.9 & 94.9\\
		& 30 & 4.49 & 4.49 & 5.00 & 5.00 & 0.97 & 0.93 & 0.94 & 0.94 &  91.3 & 91.2 & 94.7 & 94.7\\
		\hline
		\multirow{4}{*}{0.05} 
		& 5  & 4.45 & 4.43 & 4.96 & 4.97 & 1.88 & 1.76 & 1.81 & 1.80 & 94.0 & 93.7 & 95.3 & 97.1 \\
		& 10 & 4.51 & 4.49 & 5.01 & 5.01 & 1.36 & 1.28 & 1.30 & 1.29 & 93.7 & 93.4 & 95.0 & 95.5 \\
		& 20 & 4.50 & 4.50 & 5.01 & 5.01 & 0.97 & 0.91 & 0.92 & 0.92 & 91.9 & 91.6 & 94.8 & 94.8 \\
		& 30 & 4.50 & 4.50 & 5.01 & 5.01 & 0.79 & 0.75 & 0.76 & 0.75 & 90.4 & 89.8 & 94.6 & 94.6 \\
		\hline
		\multirow{4}{*}{0.001} 
		& 5  &  4.48 & 4.46 & 4.99 & 4.99 & 1.34 & 1.18 & 1.35 & 1.39 & 93.4 & 93.5 & 98.1 & 99.4 \\ 
		& 10 &  4.50 & 4.49 & 5.02 & 5.01 & 0.96 & 0.85 & 0.93 & 0.96 & 92.3 & 91.6 & 96.9 & 97.9 \\
		& 20 &  4.49 & 4.49 & 5.00 & 5.00 & 0.69 & 0.61 & 0.65 & 0.66 & 88.9 & 87.2 & 96.3 & 96.8 \\
		& 30 &  4.48 & 4.48 & 4.99 & 4.99 & 0.56 & 0.50 & 0.52 & 0.54 & 84.9 & 81.6 & 95.6 & 96.3 \\
		\bottomrule
	\end{tabular}}
	\label{tab:case3}
\end{table}
\noindent
allow for different covariate effects in the two intervention groups (scenario 3 and 4).
  
The LMM and MI had similar empirical SEs (results not presented) of the intervention effect estimates. But the MI gave slightly overestimation of SEs for small values of $ \rho $ and   $ k $, which leads to a decrease in power. This agrees with the results for variance estimates of group means in previous studies \cite{Andridge2011}. The LMM gave coverage probabilities close to nominal level except very small $ \rho $ and small $ k $, where it showed slightly overcoverage.
\begin{table}[H]
	\caption{Simulation results-the two intervention groups have different missingness mechanisms and different covariate effects. Empirical average estimates of intervention effect, average estimated SEs and coverage probabilities of nominal 95\% interval over 10000 simulation runs using unadjusted cluster-level analysis (Uadj), baseline covariate adjusted cluster-level analysis (Adj), linear mixed model (LMM), using CCA, and multiple imputation (MI). Monte-Carlo errors for average estimates and average estimated SEs are all less than 0.025 and 0.018, respectively. The true value of the intervention effect is 5.} 
	\centering
	\scalebox{0.85}{
	\begin{tabular}{c|c|cccc|cccc|cccc}
		\toprule
		\multirow{2}{*}{$ \rho $} &	\multirow{2}{*}{$ k $}	& \multicolumn{4}{c|}{Average Estimate}  & \multicolumn{4}{c|}{Average estimated SE} & \multicolumn{4}{c}{Coverage (\%)}\\
		\cline{3-14}
		&	& Uadj & Adj  & LMM & MI & Uadj & Adj &  LMM & MI & Uadj & Adj &  LMM & MI \\
		\midrule
		\multirow{4}{*}{0.1} 
		& 5  & 3.02 & 4.09 & 5.00 & 5.00 & 2.44 & 2.31 & 2.42 & 2.37 & 89.0 & 93.4 & 95.7 & 98.1\\
		& 10 & 3.03 & 4.10 & 5.01 & 5.01 & 1.76 & 1.67 & 1.73 & 1.71 & 82.0 & 93.5 & 95.8 & 96.3\\
		& 20 & 3.03 & 4.11 & 5.01 & 5.01 & 1.25 & 1.19 & 1.23 & 1.23 & 66.6 & 88.8 & 95.6 & 95.6\\
		& 30 & 3.03 & 4.11 & 5.01 & 5.02 & 1.02 & 0.97 & 1.01 & 1.01 & 52.8 & 85.9 & 95.2 & 95.2\\
		\hline
		\multirow{4}{*}{0.05} 
		& 5  & 3.02 & 4.10 & 5.01 & 5.01 & 2.05 & 1.89 & 2.06 & 2.04 & 87.0 & 93.9 & 96.5 & 99.0\\
		& 10 & 3.02 & 4.10 & 5.01 & 5.01 & 1.47 & 1.36 & 1.45 & 1.44 & 75.9 & 90.4 & 95.7 & 96.7\\
		& 20 & 3.01 & 4.08 & 4.98 & 4.98 & 1.05 & 0.98 & 1.03 & 1.03 & 55.3 & 84.9 & 95.8 & 95.9\\
		& 30 & 3.02 & 4.10 & 5.01 & 5.00 & 0.86 & 0.80 & 0.84 & 0.84 & 38.0 & 81.1 & 95.6 & 95.7\\ 
		\hline
		\multirow{4}{*}{0.001} 
		& 5  & 3.02 & 4.07 & 4.99 & 4.99 & 1.57 & 1.37 & 1.69 & 1.75 & 80.4 & 91.1 & 98.5 & 99.8 \\
		& 10 & 3.03 & 4.10 & 5.00 & 5.00 & 1.13 & 0.99 & 1.17 & 1.21 & 63.0 & 87.6 & 97.6 & 98.7 \\
		& 20 & 3.02 & 4.10 & 5.00 & 5.00 & 0.81 & 0.71 & 0.81 & 0.84 & 33.4 & 77.7 & 97.0 & 97.7 \\
		& 30 & 3.01 & 4.10 & 5.00 & 5.00 & 0.66 & 0.58 & 0.66 & 0.68 & 16.7 & 67.9 & 96.5 & 97.1 \\   
		\bottomrule
	\end{tabular}}
	\label{tab:case4}
\end{table}
The average estimates of adjusted DF $ (\nu_\text{adj}) $, used by MI, over 10000 simulations runs and the calculated complete data DF $ (\nu_\text{com}) $ for scenario 4 are presented in Table \ref{tab:case5}.  Results showed that there was a severe underestimation of  $ \nu_\text{adj} $ compared to the $ \nu_\text{com} $, and that the underestimation is more severe for small values of $ \rho $.  This underestimation of $ \nu_\text{adj} $ with the overestimation of SEs resulted in overcoverage for MI.
\begin{table}[t]
	\caption{ Comparison between the complete data DF $ (\nu_{\text{com}}) $ and the average estimates of adjusted DF $ ( \nu_\text{adj}) $, over 10000 simulation runs, used by MI, when the two intervention groups have different missingness mechanisms and different covariate effects. The last two columns shows the  upper 2.5\% points of the $ t- $distribution with $ \nu_{\text{com}} $ and  $ \nu_{\text{adj}} $ degrees of freedom, respectively.} 
	\centering
	\begin{tabular}{c|c|cr|cc}
		\toprule
		$ \rho $ &	$ k $  & $\nu_{\text{com}}$ & $\nu_\text{adj} $ & $ t_{\nu_{\text{com}}}(0.025) $ & $ t_{\nu_{\text{adj}}}(0.025) $ \\
		\midrule
  \multirow{4}{*}{0.1}  
  & 5  & 8  & 4.58   & 2.31 & 2.64\\
  & 10 & 18 & 11.72  & 2.10 & 2.18\\
  & 20 & 38 & 25.71  & 2.02 & 2.06\\
  & 30 & 58 & 38.74  & 2.00 & 2.02\\  
	 \hline 
	 \multirow{4}{*}{0.05}  
    & 5  & 8  & 3.92   & 2.31 & 2.80\\
	& 10 & 18 & 9.64   & 2.10 & 2.24\\
	& 20 & 38 & 20.61  & 2.02 & 2.08\\
    & 30 & 58 & 30.18  & 2.00 & 2.04\\
	 \hline  
	\multirow{4}{*}{0.001}  
    & 5  & 8  & 3.12  & 2.31 & 3.11\\
    & 10 & 18 & 7.12  & 2.10 & 2.36 \\
    & 20 & 38 & 13.73 & 2.02 & 2.14\\
    & 30 & 58 & 19.01 & 2.00 & 2.09\\
	 \bottomrule                           
    \end{tabular}
	\label{tab:case5}
\end{table}
\section{Discussion and Conclusion}
\label{discussion_conclusion}
Cluster randomised trials have increasingly being accepted to evaluate interventions  which are usually applied at the cluster level instead of individual level in health service research. The risk of attrition in such trials might be high  due to lack of direct contact with participants and often lengthy follow-up period \cite{donner1990}. In such trials, a plausible working assumption, particularly if baseline covariates are collected which are related to missingness, that missingness in outcome depends on covariates measured at baseline, but not on the outcome itself, known as covariate dependent missingness. Adjustment for baseline covariates in CRTs can increase power to detect a intervention effect. Cluster-level analysis is widely used by applied researchers because of its simplicity.  In this paper, we focused on selection of an analysis method for CRTs where outcome (continuous) is missing under covariate dependent missingness.  We found that unadjusted cluster-level analysis and baseline covariate adjusted cluster-level analysis give unbiased estimates of intervention effect only if both intervention groups have the same missingness mechanisms and the same covariate effects, which is a very strong assumption about missingness in CRTs and is arguably unlikely to hold in practice. We therefore caution researchers that these methods may commonly give biased inferences in CRTs suffering from missingness in outcomes.  The LMM and MI gave unbiased estimates of intervention effect regardless of whether missingness mechanisms and covariate effects are same or different in two groups, provided that an interaction between intervention indicator and covariate was included in the model. According to Groenwold \cite{Groenwold2012}, there is little to be gained by using MI over LMM in the absence of auxiliary variables. This is due to the fact that when missingness is confined to outcomes, mixed models fitted using maximum likelihood are fully efficient and valid under MAR. Consequently, in the absence of auxiliary variables, LMM can be recommended as the primary analysis approach for CRTs with missing outcomes if one is willing to assume make the CDM assumption.  

Throughout this paper, we have assumed CDM mechanism in a continuous outcome, which is an example of MAR as our baseline covariate was fully observed. In practice, we can not definitely identify on the basis of the observed data  which missingness assumption is appropriate \cite{White2010,carpenterKen2013}. Therefore, in practice, ideally sensitivity analyses should be performed \cite[Ch. 10]{carpenterKen2013} to explore whether our inferences are robust to the primary working assumption regarding the missingness mechanism. Furthermore, we focused on studies with only one individual-level covariate; the methods described can be extended for more than one covariates. 

\begin{appendices}
\section{Unadjusted cluster-level analysis using CCs}
\label{appenA}
The mean of the observed outcomes in a particular cluster can be written as
\begin{eqnarray}
\bar{Y}_{ij}^{\text{obs}}
&=& \frac{1}{\sum_{l}^{m_{ij}}R_{ijl}}\sum_{l=1}^{m_{ij}} R_{ijl}Y_{ijl}\nonumber\\
&=& \frac{1}{\sum_{l}^{}R_{ijl}}\sum_{l=1}^{m_{ij}}R_{ijl}\left( \alpha_i+ \beta_i X_{ijl}+\delta_{ij}+\epsilon_{ijl}\right)\nonumber\\
&=&\alpha_i+\beta_i\frac{1}{\sum_{l}^{}R_{ijl}}\sum_{l=1}^{m_{ij}}R_{ijl}X_{ijl}+\delta_{ij}+\frac{1}{\sum_{l}^{}R_{ijl}} \sum_{l=1}^{m_{ij}} R_{ijl}\epsilon_{ijl} \nonumber\\
&=&\alpha_i+\beta_i\bar{X}_{ij}^{\text{obs}}+\delta_{ij}+\frac{1}{\sum_{l}^{}R_{ijl}} \sum_{l=1}^{m_{ij}} R_{ijl}\epsilon_{ijl} \nonumber,
\end{eqnarray}
where $ \bar{X}_{ij}^{\text{obs}}=\left( {1}/{\sum_{l}^{}R_{ijl}}\right) \sum_{l=1}^{m_{ij}}R_{ijl}X_{ijl} $ is the observed mean of the baseline covariate $ X $ in the $ (ij) $th cluster. The expected value of $ \bar{X}_{ij}^{\text{obs}}  $ across the clusters in the  $ i $th intervention group will be the true mean of $ X $ among those individuals with observed outcomes. Let $ \mu_{xi1} $ denote the true mean of the baseline covariate $ X $ in the $ i $th intervention group among those individuals with observed outcomes. Then
\begin{eqnarray}
\mbox{E}\left(\bar{Y}_{ij}^{\text{obs}} \right)= \alpha_i+\beta_i\mu_{xi1}+\mbox{E}\left(\frac{1}{\sum_{l}^{}R_{ijl}} \sum_{l=1}^{m_{ij}} R_{ijl}\epsilon_{ijl} \right) \nonumber
\end{eqnarray}
Let $ \boldsymbol{R_{ij}}=(R_{ij1},R_{ij2},\ldots,R_{ijm})$ be the vector of missing data indicators for the $ (ij) $th cluster. Then
\begin{eqnarray}
\mbox{E}\left(\frac{1}{\sum_{l}^{m_{ij}}R_{ijl}} \sum_{l=1}^{m_{ij}} R_{ijl}\epsilon_{ijl} \right)
&=& \mbox{E}\left[ \mbox{E}\left(\frac{1}{\sum_{l}^{}R_{ijl}} \sum_{l=1}^{m_{ij}} R_{ijl}\epsilon_{ijl}\Big|\boldsymbol{R_{ij}} \right)\right] \nonumber\\
&=& \mbox{E}\left[ \frac{1}{\sum_{l}^{}R_{ijl}} \sum_{l=1}^{m_{ij}} R_{ijl}\mbox{E}\left( \epsilon_{ijl}\Big|\boldsymbol{R_{ij}}\right) \right] \nonumber\\
&=& 0,
\label{exp1}
\end{eqnarray}
since $ \epsilon_{ijl} $'s are independent of $ R_{ijl} $'s and $ \mbox{E}(\epsilon_{ijl})=0 $. Therefore, we have
\begin{equation}
\mbox{E}\left(\bar{Y}_{ij}^{\text{obs}} \right)= \alpha_i+\beta_i\mu_{xi1}.\nonumber
\end{equation}
The variance of $ \bar{Y}_{ij} $ can be written as
\begin{eqnarray}
\mbox{Var}\left(\bar{Y}_{ij}^{\text{obs}} \right) & = & \beta_i^{2}\mbox{Var}\left(\bar{X}_{ij}^{\text{obs}} \right)+\sigma_b^2+\mbox{Var}\left(\frac{1}{\sum_{l}^{}R_{ijl}} \sum_{l=1}^{m_{ij}} R_{ijl}\epsilon_{ijl} \right)\nonumber\\
&=& \beta_i^{2}\sigma^2_{\bar{x}_{i1}}+\sigma_b^2+\mbox{Var}\left(\frac{1}{\sum_{l}^{}R_{ijl}} \sum_{l=1}^{m_{ij}} R_{ijl}\epsilon_{ijl} \right)\nonumber,
\end{eqnarray}
$ \sigma^2_{\bar{x}_{i1}} $ is the variance of $ \bar{x}_{i1} $, the sample mean of $ X $ values in the $ i $th intervention group among those individuals with observed outcomes.

\noindent
Now
\begin{eqnarray}
\mbox{Var}\left( \frac{1}{\sum_{l}^{m_{ij}}R_{ijl}}\sum_{l=1}^{m_{ij}}R_{ijl}\epsilon_{ijl}\right) 
& = & \mbox{Var}\left[\text{E}\left(  \frac{1}{\sum_{l}^{}R_{ijl}}\sum_{l=1}^{m_{iij}}R_{ijl}\epsilon_{ijl}\Big |\boldsymbol{R}_{ij}\right)\right]\nonumber \\
&& +\mbox{E}\left[\text{Var}\left(  \frac{1}{\sum_{l}^{}R_{ijl}}\sum_{l=1}^{m_{ij}}R_{ijl}\epsilon_{ijl}\Big |\boldsymbol{R}_{ij}\right)\right]\nonumber\\
& = & 0 +\mbox{E}\left[  \frac{1}{\left( \sum_{l}^{}R_{ijl}\right) ^{2}}\sum_{l=1}^{m_{ij}}R_{ijl}\text{Var}\left(\epsilon_{ijl}\Big |\boldsymbol{R}_{ij}\right)\right],\nonumber\text{~using ~}(\ref{exp1})\\
& = & \sigma_{w}^{2}\text{E}\left(  \frac{1}{\sum_{l}^{}R_{ijl}} \right)\nonumber\\
& = & \frac{\sigma_{w}^{2}}{\eta_i},
\label{exp2}
\end{eqnarray}
where $ \text{E}\left( 1/\left( {\sum_{l}^{m_{ij}}R_{ijl}}\right)  \right)=1/\eta_i\text{~(say)~} $. Therefore,
\begin{equation}
\mbox{Var}\left(\bar{Y}_{ij}^{\text{obs}} \right)=\beta_i^{2}\sigma^2_{\bar{x}_{i1}}+\sigma_{b}^{2}+\frac{\sigma_{w}^{2}}{\eta_i}\nonumber.
\end{equation}
The observed mean of the $ i $th intervention group is calculated  as
\begin{equation}
\bar{Y}_{i}^{\text{obs}}= \frac{1}{k_i} \sum_{j=1}^{k_i} \bar{Y}_{ij}^{\text{obs}}\nonumber
\end{equation}
Then
\begin{equation}
\mbox{E}\left( \bar{Y}_{i}^{\text{obs}}\right) = \alpha_i + \beta_i\mu_{xi1}\nonumber.
\end{equation}
and
\begin{equation}
\mbox{Var}\left(\bar{Y}_{i}^{\text{obs}} \right)=\frac{1}{k_i}\left( \beta_i^{2}\sigma^2_{\bar{x}_{i1}}+\sigma_{b}^{2}+\frac{\sigma_{w}^{2}}{\eta_i}\right). \nonumber
\end{equation}
The  estimator of intervention effect  in unadjusted cluster-level analysis based on observed values is given by
\begin{equation}
\hat{\theta}^{\text{obs}} = \bar{Y}_{1}^{\text{obs}} - \bar{Y}_{2}^{\text{obs}}\nonumber.
\end{equation}
Then
\begin{eqnarray}
\mbox{E}\left( \hat{\theta}^{\text{obs}} \right)
& = &  \left( \alpha_1+\beta_1\mu_{x11}\right) -\left( \alpha_2+\beta_2\mu_{x21}\right)  \nonumber\\
& = & \left( \alpha_1+\beta_1\mu_{x}\right) - \left( \alpha_2+\beta_2\mu_{x}\right) +\beta_1\left( \mu_{x11}-\mu_{x}\right) -\beta_2\left( \mu_{x21}-\mu_{x}\right)\nonumber\\
& = & \mu_1-\mu_2 +\beta_1\left( \mu_{x11}-\mu_{x}\right) -\beta_2\left( \mu_{x21}-\mu_{x}\right)\nonumber.
\end{eqnarray}
and
\begin{eqnarray}
\text{Var}\left(  \hat{\theta}^{\text{obs}} \right) 
& = &  \frac{1}{k_1}\left( \beta_1^{2}\sigma^2_{\bar{x}_{11}}+\sigma_{b}^{2}+\frac{\sigma_{w}^{2}}{\eta_1}\right)+\frac{1}{k_2}\left( \beta_2^{2}\sigma^2_{\bar{x}_{21}}+\sigma_{b}^{2}+\frac{\sigma_{w}^{2}}{\eta_2}\right)\nonumber\\
&=& \sum_{i=1}^{2} \frac{1}{k_i}\left( \beta_i^{2}\sigma^2_{\bar{x}_{i1}}+\sigma_{b}^{2}+\frac{\sigma_{w}^{2}}{\eta_i}\right)\nonumber, 
\end{eqnarray}
which tends to zero as $ (k_1 , k_2) $ tend to infinity.	
	
\section{Adjusted cluster-level analysis}
\label{AppenB}
The mean of observed residuals of a particular cluster is given by
\begin{eqnarray}
{\bar{\hat{\epsilon}}_{ij}}^\text{obs}  
&=&\frac{1}{\sum_{l}^{m_{ij}}R_{ijl}}\sum_{l=1}^{m_{ij}}R_{ijl} \hat{\epsilon}_{ijl}\nonumber\\
&=& \frac{1}{\sum_{l}^{}R_{ijl}}\sum_{l=1}^{m_{ij}}R_{ijl}\left( Y_{ijl}-\hat{Y}_{ijl} \right) \nonumber\\
& = & \frac{1}{\sum_{l}^{}R_{ijl}}\sum_{l=1}^{m_{ij}}R_{ijl}\left( \alpha_i+ \beta_i X_{ijl}+\delta_{ij}+\epsilon_{ijl}-\gamma-\lambda X_{ijl}  \right)\nonumber\\
&=& \alpha_i +\left( \beta_i -\lambda \right) \frac{1}{\sum_{l}^{}R_{ijl}}\sum_{l=1}^{m_{ij}}R_{ijl}X_{ijl}+\delta_{ij} +\frac{1}{\sum_{l}^{}R_{ijl}}\sum_{l=1}^{m_{ij}}R_{ijl} \epsilon_{ijl}-\gamma\nonumber\\
&=& \alpha_i +\left( \beta_i -\lambda \right) \bar{X}_{ij}^{\text{obs}} +\frac{1}{\sum_{l}^{}R_{ijl}}\sum_{l=1}^{m_{ij}}R_{ijl} \epsilon_{ijl}-\gamma\nonumber
\end{eqnarray}
Then
\begin{equation}
\mbox{E}\left({\bar{\hat{\epsilon}}_{ij}}^\text{obs}\right)=\alpha_i +\left( \beta_i -\lambda \right) \mu_{xi1}-\gamma \nonumber
\end{equation}
and
\begin{equation}
\mbox{Var}\left({\bar{\hat{\epsilon}}_{ij}}^\text{obs} \right)=\left( \beta_i-\lambda  \right)^2\sigma^2_{\bar{x}_{i1}} +\sigma_b^2+\frac{\sigma_w^2}{\eta_i}\nonumber,
\end{equation}
using the results (\ref{exp1}) and (\ref{exp2}). The mean of observed residuals of the $ i $th intervention group can be written as
\begin{equation}
{\bar{\hat{\epsilon}}_{i}}^\text{obs}= \frac{1}{k_i} \sum_{j=1}^{k_i} {\bar{\hat{\epsilon}}_{ij}}^\text{obs}\nonumber
\end{equation}
Then
\begin{equation}
\mbox{E}\left({\bar{\hat{\epsilon}}_{i}}^\text{obs}\right)= \alpha_i +\left( \beta_i -\lambda \right) \mu_{xi1}-\gamma\nonumber
\end{equation}
and
\begin{equation}
\mbox{Var}\left({\bar{\hat{\epsilon}}_{i}}^\text{obs}\right) =\frac{1}{k_i}\left( \left( \beta_i-\lambda  \right)^2\sigma^2_{\bar{x}_{i1}} +\sigma_b^2+\frac{\sigma_w^2}{\eta_i}  \right)\nonumber. 
\end{equation}
The  baseline covariate adjusted estimator of intervention effect, based on observed values, is given by
\begin{equation}
\hat{\theta}_{\text{adj}}^{\text{obs}} = {\bar{\hat{\epsilon}}_{1}}^\text{obs} - {\bar{\hat{\epsilon}}_{2}}^\text{obs}\nonumber
\end{equation}
Then
\begin{eqnarray}
\mbox{E}\left( \hat{\theta}_{\text{adj}}^{\text{obs}} \right) 
& = & \left( \alpha_1 +\left( \beta_1-\lambda \right) \mu_{x11}-\gamma\right) -\left( \alpha_2 +\left( \beta_2 -\lambda \right) \mu_{x21}-\gamma\right) \nonumber\\
& = & \left( \alpha_1+\beta_1\mu_{x}\right) - \left( \alpha_2+\beta_2\mu_{x}\right) +\beta_1\left( \mu_{x11}-\mu_{x}\right) -\beta_2\left( \mu_{x21}-\mu_{x}\right)+\lambda
\left( \mu_{x21}-\mu_{x11} \right)\nonumber\\
& = & \mu_1-\mu_2 +\beta_1\left( \mu_{x11}-\mu_{x}\right) -\beta_2\left( \mu_{x21}-\mu_{x}\right)+\lambda\left( \mu_{x21}-\mu_{x11} \right) \nonumber
\end{eqnarray}
and
\begin{eqnarray}
\mbox{Var}\left(\hat{\theta}_{\text{adj}}^{\text{obs}}  \right)
&=& \frac{1}{k_1}\left( \left( \beta_1-\lambda  \right)^2\sigma^2_{\bar{x}_{11}} +\sigma_b^2+\frac{\sigma_w^2}{\eta_1}  \right)+\frac{1}{k_2}\left( \left( \beta_2-\lambda  \right)^2\sigma^2_{\bar{x}_{21}} +\sigma_b^2+\frac{\sigma_w^2}{\eta_2}  \right)\nonumber\\
&=& \sum_{i=1}^{2} \frac{1}{k_i}\left( \left( \beta_i-\lambda  \right)^2\sigma^2_{\bar{x}_{i1}} +\sigma_b^2+\frac{\sigma_w^2}{\eta_i}  \right)\nonumber
\end{eqnarray}
which tends to zero as $ (k_1, k_2) $ tend to infinity. 
	
\end{appendices}

\end{document}